# X-RAY ABSORPTION (XANES) AND PHOTOELECTRON SPECTROSCOPY (XPS) OF GAMMA IRRADIATED Nd DOPED PHOSPHATE GLASS

V. N. Rai[a], Parasmani Rajput[b], S. N. Jha[b], D. Bhattacharya[b], B. N. Raja Shekhar[b],

[a] Indus Synchrotron Utilization Division

Raja Ramanna Centre for Advanced Technology, Indore-452013, India

[b]Atomic & Molecular Physics Division, Bhabha Atomic Research Centre, Trombay, Mumbai-400085, India

Email: vnrai@rrcat.gov.in

Phone:+91-731-2488142 (O)

## ABSTRACT

This paper presents the X-ray absorption near edge structure (XANES) and X-ray photoelectron spectroscopic (XPS) studies of Nd doped phosphate glasses before and after gamma irradiation. The intensity and location of white line peak of Nd $L_{III}$ edge are found to be dependent on the concentration of Nd as well as on the ratio of O/Nd in the glass matrix. The decrease in the peak intensity of white line of XANES spectra and asymmetry in the profile of Nd $3d_{5/2}$ peak from XPS after gamma irradiation clearly indicates that $Nd^{3+}$ gets reduced to $Nd^{2+}$ in the glass matrix, which increases with an increase in the doses of irradiation. Sharpening of Nd $3d_{5/2}$ XPS profile indicates about the deficiency of oxygen in the glass after gamma irradiation, which is supported by EDX measurement.

# 1.    Introduction

Study of rare earth doped glasses and crystals has recently generated more interest due to its importance in the development of many optoelectronic devices such as lasers, light converters, sensors, high density memories, optical fibers and laser amplifiers [1-13]. Various types of glasses such as silicate, phosphate, borate, fluoride and telluride have been used as matrix for doping trivalent rare earth ions to produce various active optical devices such as lasers and infrared to visible up-conversion phosphors, etc [1-13]. Particularly, Nd doped phosphate glasses have been widely used as laser materials [14-15] for making high power lasers, because they can accommodate large concentrations of active ions without losing the useful properties of the material. These glasses can be prepared easily with a certain range of compositional possibilities in order to facilitate tailoring of the physical, chemical and optical properties of interest for specific technological applications [16-21]. Along with above possibilities, optical properties of these glasses are also strongly dependent on the local structures near the rare earth sites as well as on the distribution of the rare earth ions in the glass matrix [22-23]. The local structural properties of glasses are determined mainly by the type, concentration and arrangement of the ligands surrounding the rare earth ions.

Study of modifications in the optical and structural properties of glasses under the effect of high energy radiation (UV, gamma rays and neutrons) has also been found to be applicable in the optics onboard space craft, in the image guides for reactor inspection, in the optical fiber guides as well as in the mobilization of high level radioactive waste [24-33]. Normally these structural changes in glasses after irradiation are found to be associated with the change in the concentration of atomic content in the glass matrix due to bond breaking and its possible reorganization. Generation of different types of defects as a result of bond breaking in glass matrix also produces corresponding changes in the optical properties of glasses [24-33]. Even electron and hole pairs generated in the glass after irradiation also play an important role in producing various changes in the properties of glasses in the form of generation of different types of defects and color centers. Particularly, the dopants such as transition metal and/or lanthanide ions capture negatively charged electrons or positively charged holes creating defects in the form of a change in their valence state as a result of photo chemical reaction during exposure of successive gamma irradiation. Zang et al. [34] have reported that $Yb^{3+}$ changes to $Yb^{2+}$ in YAG crystal after gamma irradiation, where as this process recovered after annealing of crystal. Similar changes in the valence state of $Nd^{3+}$ to $Nd^{2+}$ have been reported by Rai et al. [25, 33, 35] in neutron irradiated $Nd_2O_3$ powder and gamma irradiated Nd doped phosphate glasses. Most of these studies have been carried out using optical spectroscopic techniques, which provide information about the changes in their optical properties particularly absorption and photo luminescence. Presence of such defects in glasses and its effects on the structural properties of glasses have also been studied using various other techniques such as Fourier transform infrared spectroscopy (FTIR), Raman spectroscopy, X-ray absorption spectroscopy (XAFS) and X-ray photoelectron spectroscopy (XPS) etc. Out of these techniques X-ray absorption spectroscopy has been found to be efficient in getting information about the local environment around the doped elements in glass matrix. Particularly, X-ray absorption near edge structure (XANES) has provided information about the changes in the valence state of ions in the glass matrix after high energy beta and X-ray irradiation [36-37]. Study of structural properties of glasses has also been performed using XPS in order to find information about the valance state of dopant along with its identification. Recently, Khattak et al. [38-39] have reported the effect of laser irradiation on the local structure as well as on the valence state of copper and vanadium ions in phosphate glass using XPS technique. They have also studied the behavior of bridging and non-bridging oxygen atoms in glasses under the effect of laser irradiation.

The main aim of the present study is to find information about the structural changes in Nd doped phosphate glasses as a result of change in the composition of basic materials used for making the glasses. The effect of gamma irradiation in changing the structural properties, particularly information about the reduction of $Nd^{3+}$ in the glass as well as the behavior of bridging and non bridging oxygen is also discussed.

## 2. Experimental details

### 2.1 Method of glass preparation

The Nd doped phosphate glass samples (Obtained from CGCRI, Kolkata, India) were prepared using different compositions and combinations of $P_2O_5$, $K_2O$, BaO, $Al_2O_3$, $AlF_3$, SrO and $Nd_2O_3$ as base materials [9]. The weight percentage of each oxide taken for making four types of glass samples (sample # 1 to sample # 4) is given in table 1. Melt quenching technique was used to make these glasses, where reagents were thoroughly mixed in an agate mortar and placed in a platinum crucible for melting it in an electric furnace at $1095^0$ C for 1h 40 m. The melt was then poured onto a preheated brass plate and annealed at $365^0$ C for 18 h. Finally, the samples were polished to obtain smooth, transparent and uniform surface slab of 5 mm thickness for different measurements. The elemental compositions of each element (atomic %) in the glasses were measured using energy dispersive X-ray spectra (EDX) after glass preparation. Spectra of each sample was recorded before and after gamma irradiation using Bruker X-Flash SDD EDS detector, 129 eV in order to obtain the relative concentration of different elements present in the glass samples. Final data was obtained after averaging the data recorded at three random locations in order to compensate the variation in concentration of elements from one place to other in the same sample. Tables 2 to 4 show the average atomic % of the important elements present in the glass samples before and after gamma irradiation.

## 2.2 *Gamma irradiation of samples*

Few small pieces of glass samples were irradiated at room temperature using $^{60}$Co (2490 Ci, Gamma chamber 900) source of gamma radiation having dose rate of 2 kGy/h. Samples were irradiated for radiation doses varying from 10 to 500 kGy.

## 2.3 *XANES measurement*

Nd L$_{III}$ edge XANES spectra were measured in fluorescence mode using energy scanning EXAFS beamline (BL-9) [40-41] at Indus-2 synchrotron source (2.5 GeV, 150 mA), Raja Ramanna Centre for Advanced Technology (RRCAT), Indore, India. The beamline optics consists of a Rh/Pt coated collimating meridional cylindrical mirror followed by a Si (111) based double crystal monochromator (DCM). The second crystal of DCM is sagitally (cylindrically) bent to focus the beam in horizontal plane. Ionization chamber was used to monitor the incident beam flux $I_0$ (E), whereas a Si drift detector (VORTEX-EX) was used to detect x-ray fluorescence signal $I_f(E)$ from the samples. The total absorption coefficient μ (E) in fluorescence mode can be written as

$$\mu(E) \propto \frac{I_f(E)}{I_0(E)}$$

-------------------------------  (1)

The x-ray absorption data in the neighborhood of Nd L$_{III}$ edge (6208 eV) was recorded ranging from 6180 to 6300 eV for all the samples #1 to #4. XANES spectra of glass sample # 3 were recorded before and after gamma irradiation in order to find the effect of irradiation. Absorption spectrum of $Nd_2O_3$ was also recorded to calibrate the data obtained from glass samples.

### 2.4 XPS measurement

Core level photoelectron spectra were recorded using an OMNICRON EA125 electron spectrometer having a dual (Al and Mg) anode X-ray gun and a 125 mm concentric hemispherical electron analyzer. Al $K_\alpha$ radiation (hv = 1486.6 eV) was used as excitation source. X-ray photoelectron spectra from Nd 3d and O1s core levels were recorded using a computer controlled data collection system with the electron analyzer set at pass energy of 30 eV. The chamber pressure was maintained at $\sim 1.2 \times 10^{-9}$ Torr during this experiment. The C 1s transition at 284.6 eV was used as a reference in order to find information about the shift in spectrum due to accumulation of space charge. Normally C1s peak occurs due to hydrocarbon contamination, which is supposed to be constant irrespective of the chemical state of sample. The comparison of C 1s peaks show negligible shift in the peaks. In this experiment, XPS results have been used to find out qualitative information about the glass structure as well as to support the results obtained from XANES measurements.

### 3. Results and discussion

### 3.1 X-ray absorption spectroscopy (XANES) of glasses

The XANES spectra of Nd $L_{III}$ edge corresponding to sample #1 to #4 of Nd doped phosphate glasses are shown in Fig. 1. All the samples show white line peak, which arises from L core to bound state transition of $Nd^{3+}$ ions having electronic configuration as $1s^2 2s^2 2p^6 3s^2 3p^6 3d^{10} 4s^2 4p^6 4d^{10} 4f^3 5s^2 5p^6$. Particularly, $L_{III}$ edge white line peak occurs due to 2p $\rightarrow$ 5d ($^2P_{3/2} \rightarrow {}^2D_{5/2}$) transition [42]. The small peak after white line peak originates from multiple scattering contributions, which is known as shape resonance. Normally the white line intensity hardly shows any systematic change with rare earth atomic size, where as the second peak changes systematically due to changing bond lengths in sample material [42]. In the present experiment white line intensity obtained from glass samples has been compared based on changes in the composition of glass matrix (concentration of Nd and ratio of O/Nd) as well as after gamma irradiation of the glass samples. The white line peaks due to sample # 1 and # 3 are observed at 6209 eV where as it is at 6211 eV for sample # 2 and # 4. The peak intensities of samples # 1, # 3 and # 4 are nearly similar, where as it is less in the case of sample # 2. Except these changes the nature of spectra remains same for all the samples. The changes in the position and intensity of the white line peaks seem to be related with the concentration of Nd as well as the presence of other elements (particularly oxygen) around $Nd^{3+}$ in the glass matrix. A correlation has been observed from the data of table 2, which shows that sample # 1 and # 3 has nearly similar concentration of Nd in the glass matrix as 0.32 and 0.28 %, where as sample # 2 and # 4 has 0.09 and 0.19 % respectively. It indicates that lowest concentration of Nd in Sample # 2 may be the reason for small amplitude in the XANES spectra. Another correlation is also observed, which indicates that the peak intensity is large for lower value of O/Nd ratio in the glass samples. The shift in the peak towards higher energy seems to be related with comparatively higher value of O/Nd ratio in sample # 2 and # 4 as 914.66 and 409.84 in comparison to 243.25 and 289.50 for samples # 1 and # 3 respectively. It clearly indicates that higher value of O/Nd ratio affects the electric field around Nd in glass matrix, which ultimately affects the energy level of final $^2D_{5/2}$ state of $Nd^{3+}$ and consequently generating a slight shift in the peak position of white line towards higher energy side.

### 3.2   Effect of gamma irradiation on XANES of glasses

Effect of gamma irradiation also affects the white line intensity of XANES peak. Fig. 2 shows the Nd $L_{III}$-edge XANES spectra of $Nd_2O_3$ as well as the glass sample # 3 before and after gamma irradiation (10 and 100 kGy). This shows that $Nd_2O_3$ has white line peak located at 6210 eV. The peak due to glass sample # 3 is observed at 6211 eV having intensity lower than observed in the case of $Nd_2O_3$. This decrease in intensity is as expected due to small number density of Nd and higher value of O/Nd ratio in glass sample # 3 in comparison to $Nd_2O_3$ as discussed in previous section. Fig. 2 also shows that spectra of Nd doped phosphate glasses (before and after gamma irradiation) and $Nd_2O_3$ (Crystalline) are nearly similar. It suggests that $Nd^{3+}$ local environment in both the cases are very similar. Even the coordination geometry around $Nd^{3+}$ in the glass and that in oxide seems to be similar. Normally XANES spectra arising from the same absorbing atom in different coordination geometries are expected to be different. However observation of close similarity in the XANES spectral features of $Nd_2O_3$ and $NdF_3$-$BeF_2$ glass has also been reported earlier [42], which supports our findings. The white line peak intensity of sample # 3 shows a small decrease (Fig. 2) after gamma irradiation (10 kGy) of glass sample. The peak intensity further decreases with an increase in the doses of gamma irradiation from 10 kGy to 100 kGy. Such decrease in white line peak intensity seems to be possible only due to decrease in the number density of $Nd^{3+}$ in glass sample # 3. Similar decrease in white line intensity from XANES of $Sm^{3+}$ and $Eu^{3+}$ has been reported after X-ray irradiation of Sm and Eu doped $Na_2O$-$Al_2O_3$-$B_2O_3$ glass samples [38]. They found that $Sm^{3+}$ and $Eu^{3+}$ ions in glass samples got reduced to $Sm^{2+}$ and $Eu^{2+}$ after X-ray irradiation. In addition the pre-edge peak intensity in the XANES spectra of glasses due to $Sm^{2+}$ and $Eu^{2+}$ increases with an increase in the dose of X-ray irradiation. It clearly indicates that number density of $Sm^{3+}$ and $Eu^{3+}$ decreases and corresponding number density of $Sm^{2+}$ and $Eu^{2+}$ increases in the glass sample as a result of X-ray irradiation. In fact, presence of these pre-edge peaks in the spectra recorded from the un-irradiated glass samples indicated the existence of both types of oxidation state (as 3+ and 2+) for Sm and Eu in the $Na_2O$-$Al_2O_3$-$B_2O_3$ glass samples. In another experiment Rossano et al. [39] have reported similar variation in XANES spectra of iron bearing soda lime glass system after beta irradiation and concluded that $Fe^{3+}$ got reduced to $Fe^{2+}$ in this process. Similarly in the present experiment, decrease in the white line peak intensity due to $Nd^{3+}$ after gamma irradiation also indicates that $Nd^{3+}$ is getting reduced to $Nd^{2+}$ in phosphate glass. However no chemical shift or change in the absorption edge spectra (the spectral profile of white line) is observed as a result of change in the oxidation state of $Nd^{3+}$ to $Nd^{2+}$, which seems to be due to low concentration of induced $Nd^{2+}$ in the glass sample. These observations are in line with earlier reported results [25, 33, 35] obtained using optical spectroscopic techniques. Here we could not perform extended X-ray absorption fine structure (EXAFS) study due to low concentration of Nd in the glass samples.

### 3.3 XPS of phosphate glasses

### 3.3.1 Nd 3d spectra

X- Ray photoelectron spectra (XPS) of Nd doped phosphate glasses are recorded before and after gamma irradiation using Al $K_\alpha$ X-ray source. For this purpose X-ray is focused on the surface of glass sample. In this section, attention has been paid on the core level spectrum of Nd 3d in order to see the effect of gamma irradiation. Fig. 3 shows the XPS spectra of glass sample # 1 before and after 10 and 500 kGy gamma irradiation. The fresh sample shows two broad peaks at ~982.5 and ~1003.7 eV due to $3d_{5/2}$ and $3d_{3/2}$ (Nd 3d spectra). It is observed that these peaks become sharper in the case of gamma irradiated samples, where as peaks shift to higher binding energy side. This observation of sharpening and shift in the peaks towards higher energy side may be due to decrease in oxygen content in the glass samples after gamma irradiation. It seems that some of the oxygen peaks lying near Nd 3d peaks making the combined peak broad in the case of fresh sample. This conclusion was drawn on the basis of data obtained from EDX, which shows a decrease in the elemental concentration of the oxygen in the glass sample after gamma irradiation (Table 3 and 4). A drastic decrease in the concentration of oxygen in the glass sample #1 is noted in the case of heavy dose of irradiation (500 kGy). However decrease in the concentration of oxygen in EDX data is not observable for low dose of irradiation of 10 kGy. Inspite of this both the sample # 1 and # 3 show a sharp Nd 3d peak due to $3d_{5/2}$ after gamma irradiation at low dose of irradiation of 10 kGy (Fig. 4). It is also noted that the main peak of sample # 3 shifts to lower energy side and is broad in comparison to peak due to sample # 1, which may be due to the presence of $AlF_3$ and higher concentration of oxygen in sample # 3 network. Further the intensity of peak due to Nd $3d_{5/2}$ shows a decreasing trend after gamma irradiation (Fig. 3). The intensity of this peak is comparatively higher in the case of gamma irradiation dose of 500 kGy than 10kGy. Such changes in the intensity of peaks may be due to change in the relative concentration of elements after gamma irradiation, which is possible as a result of bond breaking and its reorganization in the glass samples due to gamma irradiation as discussed in FTIR observations reported earlier [9]. This is also possible as a result of diffusion of ions from surface to bulk and vice versa along with the emission of oxygen after gamma irradiation. Out of all these possibilities the decrease in the intensity of the Nd $3d_{5/2}$ peak seems to be contributed mainly by the decrease in the number density of $Nd^{3+}$ as a result of change in the oxidation state from $Nd^{3+}$ to $Nd^{2+}$ as discussed in previous section with the help of XANES measurement. These observations are in agreement with the optical observations reported by Rai et al. [33, 35]. It has been reported that difference optical absorption spectra of glasses (after and before gamma irradiation) show decrease in the intensity at the locations, where $Nd^{3+}$ lines are observed in the absorption spectra. This decrease is found to be dependent on the doses of gamma irradiation. Similarly photoluminescence of $Nd^{3+}$ in the glasses also decreases after gamma irradiation. These observations also indicate towards decrease in the number density of $Nd^{3+}$ in the glass after gamma irradiation as a result of change in the oxidation state of $Nd^{3+}$ to $Nd^{2+}$. Further, the profile of the XPS peak due to Nd $3d_{5/2}$ shows asymmetry towards lower energy side (Fig. 3 and 4) after gamma irradiation. This also indicates about the presence of another oxidation state of Nd in the glass samples after gamma irradiation, which seems to be $Nd^{2+}$. The correlation of above observations along with the results from XANES measurement clearly indicates that gamma irradiation induces the change in the oxidation state of $Nd^{3+}$ to $Nd^{2+}$ in the glass sample. The effect of gamma irradiation has also been observed in the form of darkening of Nd doped phosphate glasses. Similar darkening has been reported in the case of X-ray irradiated Sm and Eu doped $Na_2O$-$Al_2O_3$-$B_2O_3$ glasses. These observations also indicate that generation of photo-induced defects in the glasses is main reason behind the observation of coloring in the glasses.

As discussed in the previous section on the basis of XPS and EDX measurements that concentration of oxygen in the glass samples decreases after gamma irradiation. Similar variations in XPS of Nd containing alloy glass have also been reported by Tanaka et al [43]. They have found that the presence of higher concentration of oxygen in glass induces broadening in the peaks of $3d_{5/2}$ and $3d_{3/2}$ due to Auger O KLL and 3d satellite peaks. An O KLL peak represents the energy of the electrons ejected from the atoms due to the filling of the O 1s state (K shell) by an electron from the L shell coupled with the ejection of an electron from the L shell. Such broadening in Nd (3d) peaks due to presence of extra peaks from oxygen is not observed in the sample, where oxygen content is less. The decrease in oxygen content is accompanied by a small shift in the main peak (Nd 3d) towards higher energy. Similar decrease in oxygen concentration has been reported by Puglisi et al [44] after electron beam irradiation of glass samples. They have used XPS to study the compositional changes in the glass sample after electron beam irradiation and reported an out gassing of oxygen from the sample. It is shown in EDX measurement (Table 3 and 4) that other than oxygen each and every elements present in the glass show some change in their relative concentration after gamma irradiation. Even O/Nd ratio in the glass also decreases after gamma irradiation. It is well known that XPS and EDX are effective mainly on the surface as X-rays cannot penetrate the glass samples. Therefore, it cannot provide information about the elements present in the bulk sample. As per the above discussion the decrease in the width of XPS Nd $3d_{5/2}$ peak may be associated with the decrease in O/Nd ratio in the glass after gamma irradiation as a result of diffusion of elements from the surface to inside of the bulk material. Similar observation has been reported by Khattak et al.

[39] that concentration of Vanadium in Vanadium phosphate glass decreases after laser irradiation, which is supposed to be either due to evaporation of Vanadium from the surface of the glass or due to its diffusion from surface to bulk. Finally out of all the possibilities the decrease in the peak intensity of $3d_{5/2}$ seems mainly due to decrease in the number density of $Nd^{3+}$ as a result of reduction of $Nd^{3+}$ to $Nd^{2+}$ after gamma irradiation as has been discussed earlier on the basis of optical and XANES observations.

### 3.3.2 O 1s spectra

In most of the XPS studies of oxide glasses the O 1s spectra are more informative with respect to the structure of the glass than the cation spectra. Specifically, the binding energy of the O 1s electrons is a measure of the extent to which electrons are localized on the oxygen or in the inner-nuclear region, which is a direct consequence of the nature of bonding between the oxygen and the other cations in the glass samples [38-39]. Normally an asymmetry in the O 1s core level peaks occurs due to presence of two different types of oxygen sites in the glasses. Usually, the O 1s peaks for these glasses may arise from oxygen atoms existing in some or all of the following structural bonds: P-O-P, P-O-Nd, Nd-O-Nd, P=O and P-O-M, where M is the other cations present in the glass matrix. The oxygen atoms that are more covalently bonded to glass former atoms on both the sides are typically called bridging oxygen (BO) such as P-O-P, where as the oxygen atoms that are more ionically bonded at least on one side or double bonded to a glass former atoms are referred to as non-bridging oxygen (NBO) such as P-O-Nd, Nd-O-Nd, P=O. In this situation the binding energy of NBO remains lower than that of the BO.

The XPS O 1s spectra for sample # 1 before and after gamma irradiation (10 and 500 kGy) are shown in Fig. 5. A slight asymmetry is observed in the profile of O 1s peak obtained from fresh glass samples # 1, which also indicates about the presence of two different types of oxygen sites in this glass as discussed above. This asymmetry is observed towards higher energy side. Similar asymmetry in O 1s spectra has been reported in different types of glasses by many researchers [38-39, 45]. Normally, O 1s spectrum is fitted by two Gaussian-Lorenzian peaks in order to determine the peak position and the relative contribution of the different oxygen sites. In the present case qualitative analysis of O 1s peak asymmetry (Fig. 5) clearly indicates the presence of less number of bridging oxygen in unirradiated glass sample. This profile becomes nearly symmetric after gamma irradiation (10 and 500 kGy) probably due to the decrease in the number density of bridging oxygen as a result of bond breaking and subsequent increase in the density of non-bridging oxygen. Here an exact reason for shift in the O 1s peaks after gamma irradiation of glass is not very clear. However, it may be due to the change in the concentration of different linkages in glass matrix as a result of bond breaking after gamma irradiation. Similar asymmetry has been noted in the case of O 1s spectrum from sample # 3 as shown in Fig. 6. Here again the asymmetry decreases after gamma irradiation of the sample. These observations are in line with the results obtained from FTIR spectra of these samples indicating that gamma irradiation increases the breaking of P-O-P linkages and creating non-bridging oxygen in the form of P=O, P-O-Nd and/or P-O-M linkages [9].

## 4.    Conclusions

The $L_{III}$-edge XANES peak intensity due to $Nd^{3+}$ and its location are found mainly dependent on the concentration of Nd and the ratio of O/Nd in the glass samples respectively. Here the decrease in the XANES peak intensity of Nd in glass after gamma irradiation indicates that part of the $Nd^{3+}$ is getting reduced to $Nd^{2+}$. An observation of similarity in XANES spectra of Nd doped phosphate glass and $Nd_2O_3$ suggest that both the structures may be having similar coordination geometry around $Nd^{3+}$. The sharpening of Nd $3d_{5/2}$ peak (XPS) after gamma irradiation indicates about the deficiency of oxygen atoms in the glass matrix. The asymmetry in the XPS profile of Nd $3d_{5/2}$ peak of irradiated glass indicates  the presence of two oxidation states of Nd, which confirms the observation of XANES measurement that $Nd^{3+}$ is getting reduced to $Nd^{2+}$ in the glass samples. The comparison of XPS O 1s spectra of glass samples recorded before and after gamma irradiation shows an increase in the number of non bridging oxygen in the glass as a result of breaking and reorganization of bonds in the glass under the effect of gamma irradiation. However further investigation of the glass samples having very low concentration of Nd using XAFS is needed with an enhanced sensitivity of detector to find detailed information about the coordination number and the nearest neighbour around Nd ion.

## Acknowledgments

M. P. Kamat, HPLOL, RRCAT, Indore is thankfully acknowledged for providing glass sample for this study. Authors are grateful to S. Kher, Pragya Tiwari, D. M. Phase and A. Wadikar for providing various experimental help and discussions during this work. We also thank G. S. Lodha and N. K. Sahoo for their keen interest and encouragement in this work.

**FIGURE CAPTION**

1. Nd $L_{III}$-edge XANES spectra of Nd doped phosphate glasses. (1) Sample # 1, (2) Sample # 2, (3) Sample # 3, (4) Sample # 4

2. Nd $L_{III}$-edge XANES spectra of $Nd_2O_3$ and Nd doped glass before and after gamma irradiation. (1) Sample #3, (2) Sample # 3 irradiated at 10 kGy, (3) Sample # 3 irradiated at 100 kGy, (4) $Nd_2O_3$.

3. XPS spectra of Nd 3d from sample # 1 before and after gamma irradiation, (1) Pure sample # 1, (2) Sample # 1 irradiated at 500 kGy, (3) Sample # 1 irradiated at 10 kGy.

4. XPS spectra of Nd 3d from glass samples after gamma irradiation of 10 kGy, (1) Sample # 1,   (2) Sample # 3

5. XPS spectra of oxygen O 1s, (1) Sample #1, (2) Sample # 1 irradiated at 10 kGy, (3) Sample # 1 irradiated at 500 kGy

6. XPS spectra of oxygen O 1s, (1) Sample # 3, (2) Sample # 3 irradiated at 10 kGy

**Table 1**  Composition of materials used for making the glass samples.

| S. No. | Composition | Sam. #1 (wt. %) | Sam. #2 (wt. %) | Sam. #3 (wt. %) | Sam. #4 (wt. %) |
|---|---|---|---|---|---|
| 1. | $P_2O_5$ | 58.95 | 58.50 | 55.50 | 58.50 |
| 2. | $K_2O$ | 17.00 | 17.00 | 14.00 | 17.00 |
| 3. | BaO | 14.95 | 14.50 | 14.50 | ---- |
| 4. | $Al_2O_3$ | 9.00 | 9.00 | 9.00 | 9.00 |
| 5. | $Nd_2O_3$ | 0.10 | 1.00 | 1.00 | 1.00 |
| 6. | $AlF_3$ | ----- | ----- | 6.00 | ----- |
| **7.** | SrO | ----- | ----- | ----- | 14.50 |

**Table 2**  Elemental composition (Atomic %) and O/Nd ratio present in the glass samples (EDX data)

| S. No. | Atomic Component | Sam. #1 (At. %) | Sam. #2 (At. %) | Sam. #3 (At. %) | Sam.#4 (At. %) |
|---|---|---|---|---|---|
| 1. | O | 77.84±6.35 | 82.32±5.50 | 81.06±5.60 | 77.87±7.47 |
| 2. | P | 13.20±0.60 | 10.27±0.43 | 9.99±0.43 | 13.64±0.73 |
| 3. | Ba | 1.94±0.45 | 1.32±0.26 | 1.77±0.37 | __________ |
| 4. | K | 3.96±0.25 | 3.58±0.16 | 4.73±0.20 | 3.19±0.20 |
| 5. | Al | 2.27±0.10 | 2.00±0.13 | 1.97±0.10 | 3.99±0.30 |
| 6. | Nd | 0.32±0.13 | 0.09±0.10 | 0.28±0.10 | 0.19±0.10 |
| 7. | O/Nd | 243.25 | 914.67 | 289.5 | 409.84 |

**Table  3** Elemental composition (Atomic %) and O/Nd ratio present in Sample #1 before and
after gamma irradiation  (EDX data)

| At. Comp. | O | P | Ba | K | Al | Nd | O/Nd |
|---|---|---|---|---|---|---|---|
| Av. At. % | 77.84 | 13.20 | 1.94 | 3.96 | 2.27 | 0.32 | 243.25 |
|  | ±6.35 | ±0.60 | ±0.45 | ±0.25 | ±0.10 | ±0.13 |  |
| Av. At. % | 75.45 | 12.86 | 2.38 | 5.27 | 3.08 | 0.30 | 251.50 |
| (γ irradiated 10 kGy) | ±4.40 | ±0.50 | ±0.30 | ±0.2 | ±0.10 | ±0.10 |  |
| Av. At. Wt. % | 66.19 | 17.00 | 3.12 | 5.90 | 3.99 | 0.37 | 178.89 |
| (γ irradiated 500 kGy) | ±4.3 | ±0.60 | ±0.45 | ±0.25 | ±0.15 | ±0.10 |  |

**Table  4**   Elemental composition (Atomic %) and O/Nd ratio present in the sample #3 before
and after gamma  irradiation  (EDX data)

| At. Component | O | P | Ba | K | Al | Nd | O/Nd |
|---|---|---|---|---|---|---|---|
| Av.At.% | 81.06 | 9.99 | 1.77 | 4.73 | 1.97 | 0.28 | 289.5 |
|  | ±5.60 | ±0.43 | ±0.37 | ±0.20 | ±0.10 | ±0.10 |  |
| Av. At. % | 82.83 | 7.21 | 0.49 | 0.99 | 3.85 | 0.34 | 243.62 |
| (γ irradiated 10 kGy) | ±8.40 | ±0.40 | ±0.20 | ±0.10 | ±0.25 | ±0.15 |  |

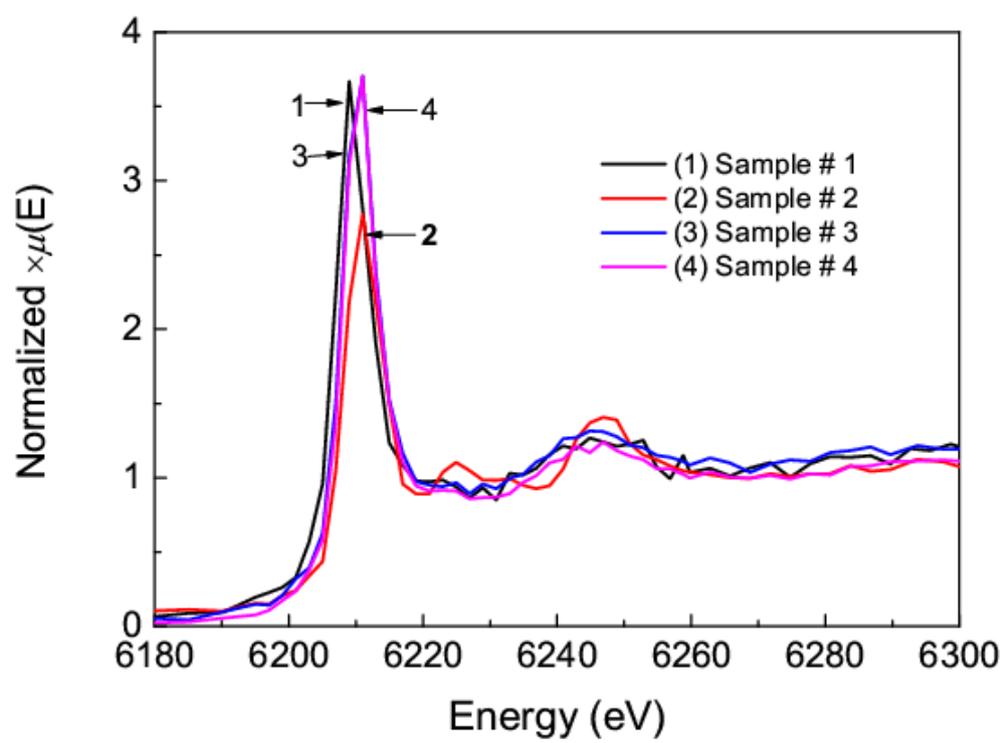

**Fig. 1**

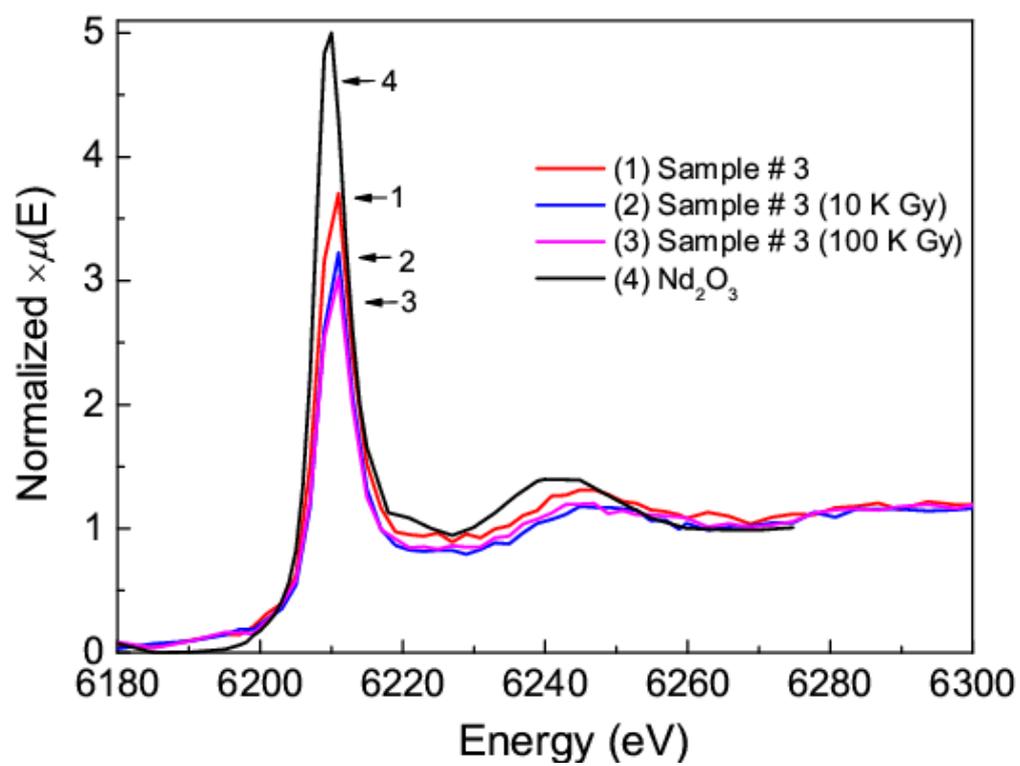

**Fig. 2**.

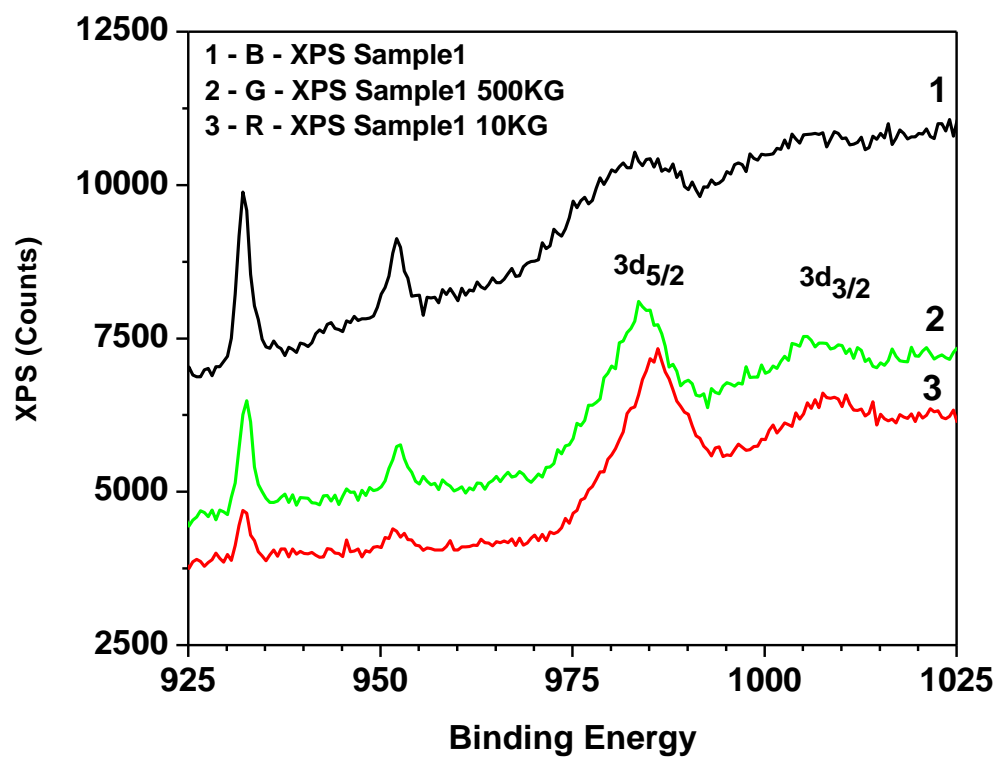

1 - B - XPS Sample1
2 - G - XPS Sample1 500KG
3 - R - XPS Sample1 10KG

**Fig. 3**

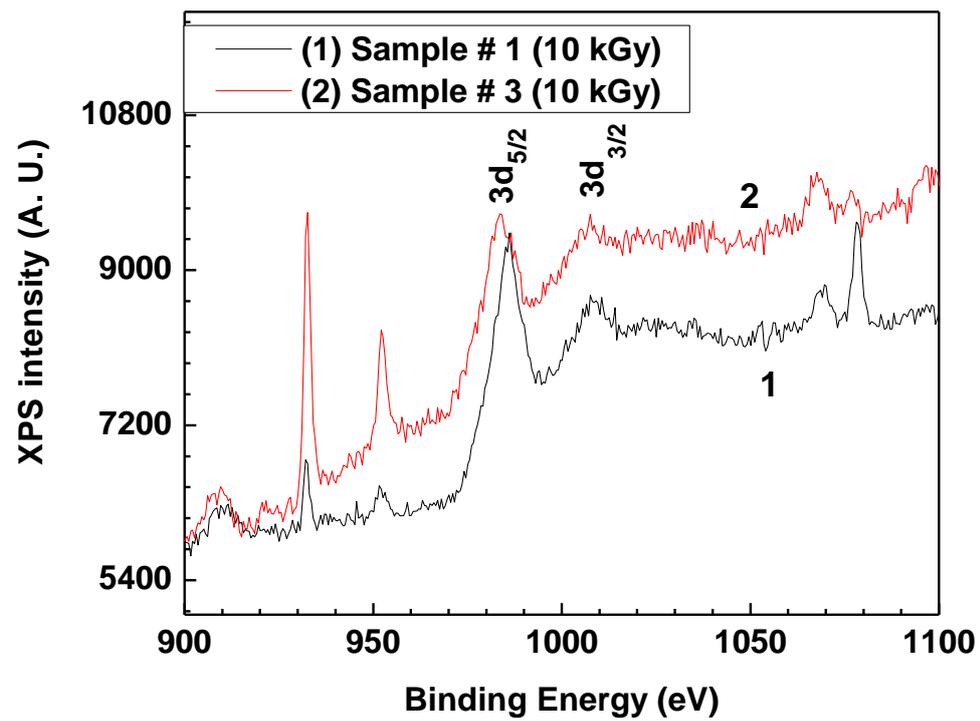

**Fig. 4**

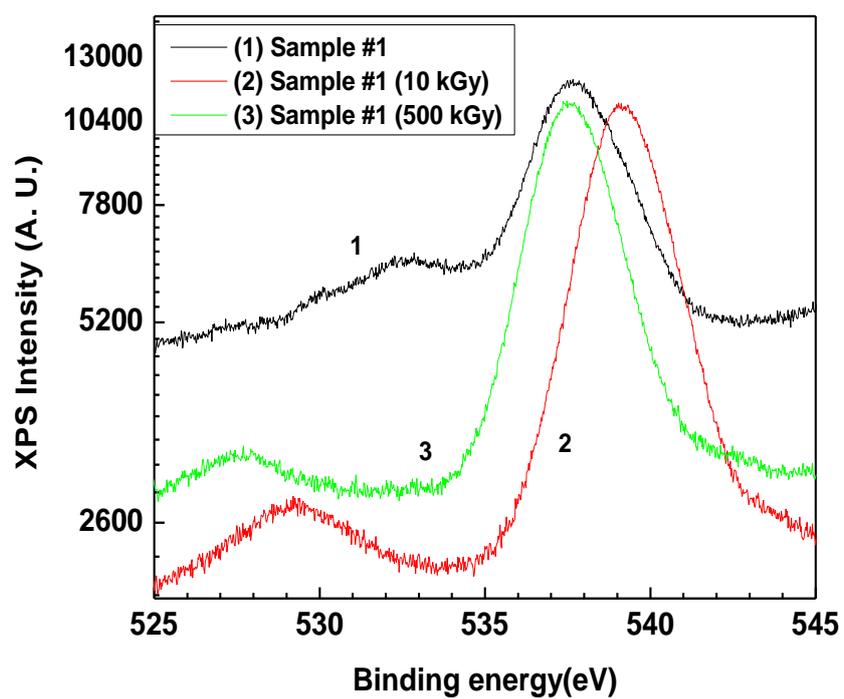

**Fig. 5**

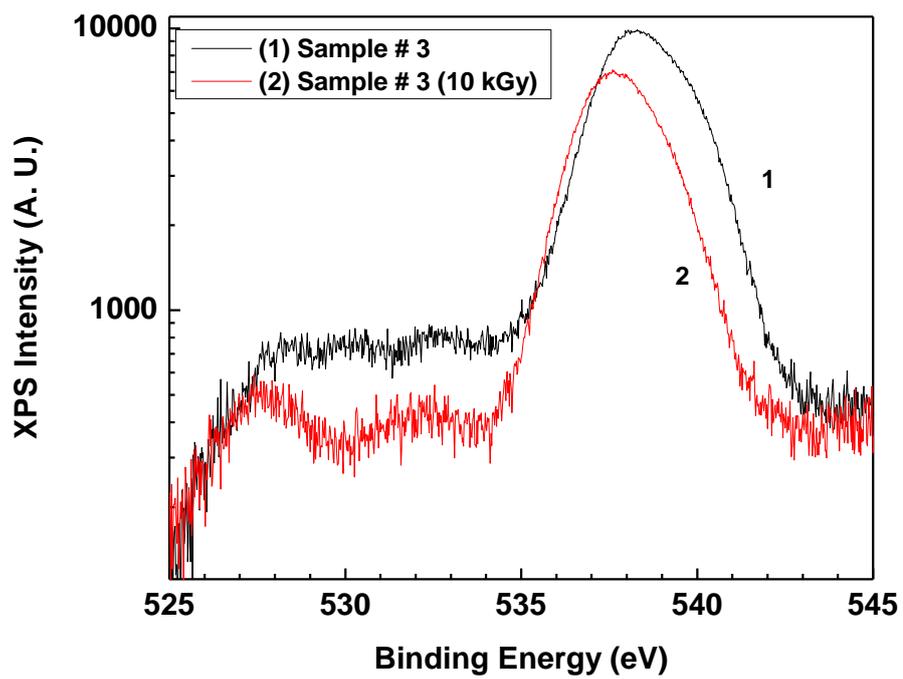

**Fig. 6**